\documentclass[review]{elsarticle}
\usepackage{graphicx}
\usepackage{epsfig,graphics,subfigure,psfrag,amsmath,amssymb}
\usepackage{lineno,hyperref}
\modulolinenumbers[5]

\begin{document}

\begin{frontmatter}

\title{Free electron nanolaser based on the graphene plasmons}

\author{H.K. Avetissian}
\author{ B.R. Avchyan}
\author{ G.F. Mkrtchian\corref{mycorrespondingauthor}}
\cortext[mycorrespondingauthor]{Corresponding author}
\ead{mkrtchian@ysu.am}
\address{Centre of Strong Fields Physics, Yerevan State University, 0025 Yerevan, Armenia}

\author{H.H. Matevosyan}
\address{Institute of Radiophysics and Electronics NAS RA, 1 Alikhanian brs.,
Ashtarak 0203, Armenia}

\begin{abstract}
In this paper, a possible way to achieve lasing from THz to extreme UV
domain due to stimulated scattering of graphene plasmons on the free
electrons is considered. The analytical-quantitative description of the
proposed FEL scheme is based on the self-consistent set of the
Maxwell--Vlasov equations. We study the downconversion as well as the
upconversion. It is shown that the coherent downconversion of infrared
radiation to THz one can be achieved using a source of very non-relativistic
electrons at the resonant coupling with the graphene plasmons. Due to the
strongly confined graphene plasmons, the upconversion of mid-infrared to
extreme UV radiation can be achieved with the mildly relativistic electron
beams. The latter is a promising mechanism for the tabletop short-wavelength
free electron nanolaser.
\end{abstract}


\end{frontmatter}

%\linenumbers

\section{ Introduction}

The free electron lasers (FEL) \cite{FEL1,FEL2,FEL3,Avetissian-Book1,Avetissian-Book2} have a
principal advantage with respect to traditional quantum generators operating
on discrete transitions that the output radiation frequency can be
continuously shifted via change of electron beam energy or stipulating
electromagnetic (EM) field parameters. The EM fields, in a general concept,
can be unified as, so called, photonic quasiparticles or quasiphotons. The
photonic quasiparticles correspond to solutions of the Maxwell's equations
in a medium with the different boundary conditions \cite{Ginzburg-Book}.
These are photons in a vacuum, medium photons, plasmons, polaritons \cite%
{Agranovich} and their confined (cavity photons, Bloch photons) \cite%
{cavity,Yab} or surface versions (surface plasmon polaritons) \cite%
{Agranovich,surface-plasmons}. All these photonic quasiparticles can be used
for the realization of FEL. The dispersion of these quasiparticles being
different from photons in a vacuum, enables many phenomena \cite%
{Avetissian-Book1,Avetissian-Book2} that impossible to realize with the photons in free space.
This primarily refers to quantum electrodynamic (QED) phenomena of the first
order. The stimulated Cherenkov effect \cite%
{Avetissian-Book1,1972Avet1,1972Avet2,ABHM1,ABHM2} has been among the first
FEL schemes with the advent of the FEL concept. From the QED perspective, it
is a first-order process and is described by the first order Feynman diagram
(that is impossible in a vacuum). Hence, the stimulated Cherenkov effect
offers the most direct and fast energy conversion from charged particles to
medium photons. However, for final lasing one needs ultra-relativistic
electron beams of high densities, which practically restricts the operation
of the Cherenkov laser in a dielectric medium due to the ionization losses.
To overcome these problems one can realize surface Cherenkov laser \cite%
{Walsh, AHM}. Another first-order process is the Smith-Purcell radiation 
\cite{SP}--emission of Bloch photons by a free-electron when it passes over
the metallic grating. The Smith-Purcell radiation can be a basic mechanism
for a compact FEL \cite{Kim,Brau,Mkr}. The second-order QED processes
represent another huge class of phenomena that can be used for FEL. These
FEL schemes rely on the scattering of photonic quasiparticles into the
output radiation by collision with the free electrons. The remarkable
examples are realized in the undulator x-ray FEL \cite{Emma,FEL3} or
proposed Compton backscattering x-ray FEL \cite{AM} schemes in the
self-amplified spontaneous emission regime, where photons energies can be
continuously upshifted up to $\gamma $-ray frequencies.

With the advent of graphene \cite{Nov} and other novel nanomaterials \cite%
{Geim}, it becomes clear that these materials support surface plasmons \cite%
{GP1,GP2,GP3,GP4} with an extreme field confinement \cite{GPc1,GPc2,GPc3}.
There is a continuos advance in increasing the confinement and lifetime of
graphene plasmons (GPs) \cite{GP4}. The confinement of\textit{\ }GPs also
means slowing down the surface EM wave's phase velocity by the same order.
The confinement and slowing down of GPs to over $200$ times smaller than the
free space parameters have been predicted \cite{GP3}. The GP lifetime
strongly depends on the quality of graphene samples which are continuously
improved with advances in fabrication techniques. The similar progress is
also seen for GPs of higher frequencies, with reports in the near-infrared
region \cite{Zhang}. This high confinement of GPs and long lifetime can be
used for the implementation of nano-FEL sources. The Cherenkov mechanism of
the FEL - electron synchronism with GPs in graphene and carbon nanotubes has
been investigated in Refs. \cite{Bat1,Bat2,Bat3,Bat4,Zhao}. The generation
and amplification of GPs by optical pumping have also been investigated in
the literature \cite{Ryzhii1,Ryzhii2,Dubinov}. All these first-order
processes have a serious drawback since it is not possible to transform
graphene GPs into free radiation because of the wavelength mismatch. This
drawback can be overcome by the various periodic constructions \cite{Liu,Li}
or by the second-order processes, where a scattering of GPs on free
electrons gives rise to output radiation (free photons) \cite{Wong,Shentcis}%
. In this case, the high momentum GPs scattering into x-ray photons using
electrons from conventional radiofrequency electron guns is possible. Hence,
it is of interest to consider the stimulated scattering of graphene plasmons
on free electrons.

In this paper, a possible way to achieve lasing from THz to extreme UV
domain due to the stimulated scattering of GPs on free electrons is
investigated. The consideration is based on the self-consistent set of the
Maxwell--Vlasov equations. We consider downconversion as well as
upconversion. It is shown that the coherent downconversion of infrared to
THz radiation can be achieved using a source of very non-relativistic
electrons ($\sim 5-25\ \mathrm{eV}$) at the resonant coupling with GPs. Due
to strongly confined GPs, the upconversion of mid-infrared to extreme UV
radiation can be achieved with mildly relativistic electron beams ($\sim 5\ 
\mathrm{MeV}$).

Our motivation is also conditioned by the recent experimental advances \cite%
{QM1,QM2,QM3} regarding the electron beam quantum modulation \cite%
{Avetissian-Book1,Avetissian-Book2,1973Avet,1976Avet,1977Avet,AM2002} using femtosecond-pulsed
lasers. Such modulated beams can serve as a source of short-wave radiation
of superradiant nature, which is also considered in the current paper.

The paper is organized as follows. In Sec. II the model and self-consistent
set of equations are presented. In Sec. III, we present a numerical solution
of the obtained equations. Finally, conclusions are given in Sec. IV.

\section{The model and self-consistent set of equations}

The configuration of the graphene-plasmon based FEL is presented in Fig. 1.
We assume a graphene layer ($x=0$) on the top of the dielectric layer of
permittivity $\varepsilon $. The dielectric layer extends from $x=-\infty $
to $x=0$. The electron beam moves in the $z$-direction and interacts with
the graphene-plasmon field. Depending on the plasmon propagation direction
we will have frequency upconversion (counterpropagating electron and
plasmon) and downconversion (copropagating electron and plasmon). We also
show a second-order Feynman diagram in which free electrons interact with
the graphene-plasmon producing outgoing photons. Notice that utmost
downconversion occurs when the mean velocity of an electron beam matches the
graphene-plasmon phase velocity. As we will see, in this case the
electron-plasmon coupling is resonantly enhancing. For a graphene, we assume
low-temperature ($T$) and high-doping limit $E_{F}>>k_{B}T$. The Fermi
energy is given as $E_{F}=\hbar \mathrm{v}_{F}\sqrt{\pi n_{s}}$, where $%
n_{s} $ is the surface carrier density, $\hbar $ is the Planck's constant,
and $\mathrm{v}_{F}$ is the Fermi velocity. In this limit, we can
approximate the conductivity as \cite{Peres}%
\begin{equation}
\sigma _{g}\left( \omega \right) =\frac{e^{2}}{\pi \hbar }\left[ \frac{iE_{F}%
}{\hbar \omega +i\hbar \gamma _{r}}+\frac{\pi }{4}\left\{ \Theta \left(
\hbar \omega -2E_{F}\right) -\frac{i}{\pi }\ln \left\vert \frac{2E_{F}+\hbar
\omega }{2E_{F}-\hbar \omega }\right\vert \right\} \right] ,  \label{sg}
\end{equation}%
where $e$ is the elementary charge, $\gamma _{r}$ is the relaxation rate, $%
\Theta $ is the Heaviside step function. The first term in the above
expression is the Drude contribution describing the intraband processes. The
second term describes interband transitions. For a heavily doped graphene,
and at the low frequencies $\hbar \omega <<2E_{F}$, the optical response is
dominated by the Drude term. The latter is the most commonly used model
describing the graphene conductivity for GPs.

For the transverse-magnetic (TM) waves we have the following dispersion
relation \cite{Bludov}: 
\begin{equation}
\frac{1}{\kappa }+\frac{\varepsilon }{\kappa _{1}}+i\frac{4\pi \sigma _{g}}{%
\omega }=0,  \label{dr}
\end{equation}%
where 
\begin{equation}
\kappa =\sqrt{q^{2}-\frac{\omega ^{2}}{c^{2}}},\ \kappa _{1}=\sqrt{q^{2}-%
\frac{\varepsilon \omega ^{2}}{c^{2}}},  \label{k1k2}
\end{equation}%
and $q$ is the complex propagation constant. 
\begin{figure}[tbp]
\includegraphics[width=.96\textwidth]{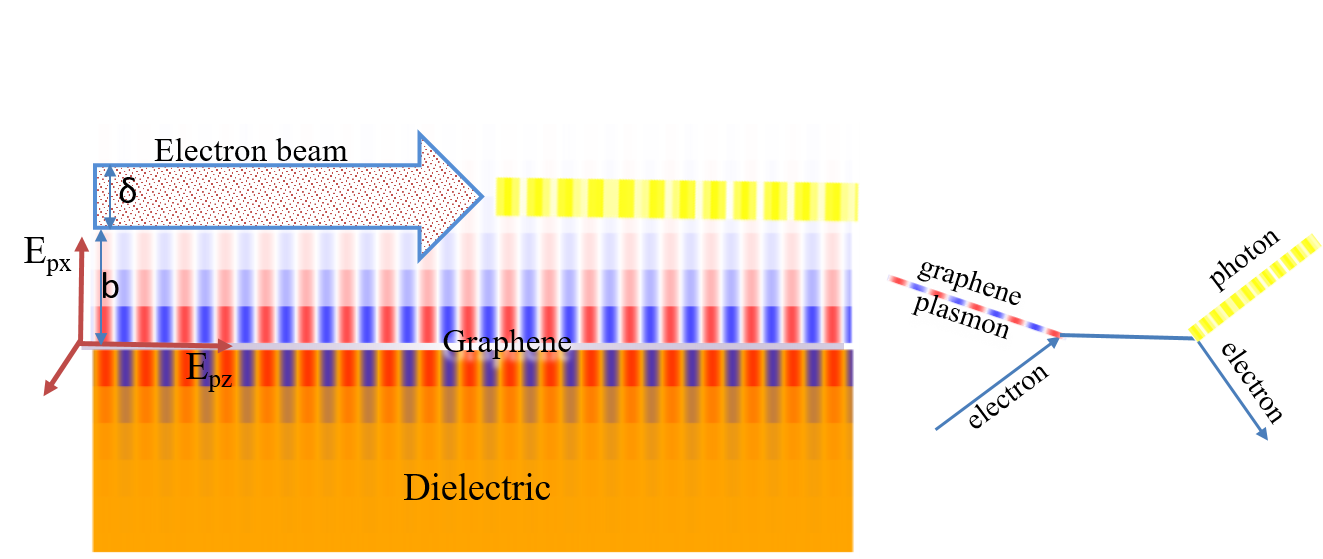}
\caption{The configuration of the graphene-plasmon based FEL. The electron
beam moves in the $z$-direction at a distance $b$ parallel to the graphene
layer which is deposited on the dielectric surface infinite in the $y$%
-direction. The electron beam thickness is $\protect\delta $. The graphene
surface is at $x=0$. Electrons interact with the graphene-plasmon field
(exponentially falling in the $\pm x$ directions red and blue peaks). On the
right, we show a Feynman diagram in which free electrons interact with the
graphene-plasmon producing outgoing photons. }
\end{figure}
The spatial confinement factor is defined as $n=c\mathrm{Re}\left( q\right)
/\omega $, where $c$ is the light speed in a vacuum. Parameter $n$
represents the degree of spatial confinement that results from the
plasmon-polariton coupling. In the limit of a large confinement factor $n>>1$
the dispersion relation can be represented as%
\begin{equation}
q=\frac{1+\varepsilon }{4\pi \sigma _{g}\left( \omega \right) }i\omega
\label{drn}
\end{equation}%
with the solution for the real ($q^{\prime }$) and imaginary ($q^{\prime
\prime }$) parts via $\sigma _{g}=\sigma _{g}^{\prime }+i\sigma _{g}^{\prime
\prime }$: 
\begin{equation*}
q^{\prime }=\frac{1+\varepsilon }{4\pi }\omega \frac{\sigma _{g}^{\prime
\prime }}{\left\vert \sigma _{g}\right\vert ^{2}},\ q^{\prime \prime }=\frac{%
1+\varepsilon }{4\pi }\omega \frac{\sigma _{g}^{\prime }}{\left\vert \sigma
_{g}\right\vert ^{2}}.
\end{equation*}%
The decay of the GPs as it propagates in space is characterized by $%
q^{\prime \prime }$. We assume that the GPs are generated simultaneously
along the entire graphene surface and $q^{\prime \prime }<<q^{\prime }$. The
confinement of the EM field in the x-direction is given by $\kappa ^{-1}$.

At first, let us consider the dynamics of the electron beam in the TM mode ($%
E_{py}=H_{px}=H_{pz}=0$) of GP wave: 
\begin{equation}
E_{px}=E_{0p}e^{-\kappa x}\sin \left( qz+s\omega _{0}t\right) ,  \label{F1}
\end{equation}%
\begin{equation}
E_{pz}=-\frac{E_{0p}\kappa }{q}e^{-\kappa x}\cos \left( qz+s\omega
_{0}t\right) ,  \label{F2}
\end{equation}%
\begin{equation}
H_{py}=-\frac{s\omega _{0}}{cq}E_{px},  \label{F3}
\end{equation}%
\qquad where $E_{0p}$ is the electric field amplitude on the graphene sheet.
The parameter $s$ has been introduced with $s=-1$ for copropagating GP wave
field and $s=1$ for counterpropagating one. Note that GP waves can be
excited using various gratings \cite{25} or with all-optical plasmon
generation at the double Fermi-edge resonances \cite{AMBM}.

The dynamics is considered in the scope of relativistic classical theory
neglecting the effects of quantum recoil, which is justified for considering
frequencies. The relativistic fluid equation for the velocity field $\mathbf{%
v}$ in the fields (\ref{F1}-\ref{F3}) reads: 
\begin{equation}
\left( \frac{\partial }{\partial t}+\mathbf{v\cdot \triangledown }\right)
\left( \mathbf{v}\gamma \right) =\frac{e}{m}\mathbf{E}_{p}+\frac{e}{mc}\left[
\mathbf{v\times H}_{p}\right] ,  \label{RH}
\end{equation}%
where $\gamma =1/\sqrt{1-\mathrm{v}^{2}/c^{2}}$ is the Lorentz factor. We
will assume the longitudinal velocity modulation is negligible, so $\mathrm{v%
}_{z}\simeq \mathrm{v}_{0}$. The transverse velocity perturbation is given
by the expression 
\begin{equation}
\mathrm{v}_{x}=\frac{eE_{0p}}{m\gamma _{0}\omega _{0}}\frac{1\mathbf{+}\frac{%
s\mathrm{\beta }_{0}}{n\left( \omega _{0}\right) }}{s+n\left( \omega
_{0}\right) \mathrm{\beta }_{0}}e^{-\kappa x}\cos \left( qz+s\omega
_{0}t\right) ,  \label{Tv}
\end{equation}%
where $\mathrm{\beta }_{0}=\mathrm{v}_{0}/c$ and $\gamma _{0}=1/\sqrt{1-%
\mathrm{\beta }_{0}^{2}}$. We will consider the case of
amplification/generation of forward radiation of the electrons. The probe EM
wave can be treated to be linearly polarized with the central frequency $%
\omega $, wave vector $k$, and electric field strength%
\begin{equation}
E_{x}=E_{0}\left( z\right) e^{i\left( kz-\omega t\right) }+\mathrm{c.c.},
\label{ez}
\end{equation}%
where $E_{0}\left( z\right) $ is a slowly varying envelope along the
electron beam propagation direction. The rate of the electron energy change
is given by the expression%
\begin{equation*}
\frac{d\mathcal{E}}{dz}=eE_{z}+\frac{eE_{x}\mathrm{v}_{x}}{\mathrm{v}_{z}},
\end{equation*}%
where $E_{z}$ is the space-charge field. Taking into account Eq. (\ref{Tv}),
we obtain 
\begin{equation}
\frac{d\mathcal{E}}{dz}=eE_{z}\mathbf{+}e\frac{\theta _{s}}{2}e^{-\kappa
x}Ee^{i\psi }+\mathrm{c.c.,}  \label{enz}
\end{equation}%
where 
\begin{equation}
\theta _{s}=\frac{eE_{0p}}{m\gamma _{0}\mathrm{v}_{0}\omega _{0}}\frac{1%
\mathbf{+}\frac{s\mathrm{\beta }_{0}}{n\left( \omega _{0}\right) }}{%
s+n\left( \omega _{0}\right) \mathrm{\beta }_{0}}  \label{intpar}
\end{equation}%
is the interaction parameter and the phase $\psi =kz-\omega t+sqz+\omega
_{0}t$. The phase $\psi $ has a following physical interpretation. For the
effective energy exchange between the electron and the wave, the phase $\psi 
$ should be kept constant along the entire interaction length $d\psi /dz=0$,
that results the condition 
\begin{equation}
\omega =\omega _{0}\frac{1+sn\left( \omega _{0}\right) \mathrm{\beta }_{0}}{%
1-\mathrm{\beta }_{0}},  \label{res_freq}
\end{equation}%
which represents the general resonance condition for the forward radiation.
As is seen from Eq. (\ref{res_freq}), for copropagating GP ($s=-1$) the
utmost downconversion occurs when the mean velocity of an electron beam
matches the graphene-plasmon phase velocity. In this case, the
electron-plasmon coupling (\ref{intpar}) is resonantly enhancing.

The transverse velocity perturbations are small enough, hence we will
consider the transverse coordinate $x$ as a parameter but not a dynamical
variable. The initial electron beam density we assume: 
\begin{equation}
n_{0}(x)=\left\{ 
\begin{array}{c}
0,\quad x<b \\ 
n_{0},\quad b\leq x\leq b+\delta \\ 
0,\quad x>b+\delta%
\end{array}%
\right. .  \label{n0}
\end{equation}%
Thus, when the beam width $\delta $ is much larger than the confinement of
the EM field in the $x$-direction: $\delta >>\kappa ^{-1}$, then one can
replace the evanescent wave factor $e^{-\kappa x}$ in the interaction
parameter by $e^{-\kappa b}$ \cite{Kim,Mkr}. At that, we can consider the
FEL theory to be one-dimensional. As in the conventional FEL theory for the
kinetic equations, it is convenient to change the independent variables from
($z,t$) to ($z,\psi $). The conjugate variable to $\psi $ will be $\mathcal{P%
}=$ $\mathcal{E-E}_{0}$, where $\mathcal{E}_{0}$ is the resonant energy
defined from Eq. (\ref{res_freq}). Thus, the effective Hamiltonian will be 
\begin{equation}
H=\mathcal{P}\Delta +\left( \omega -\omega _{0}\right) \frac{1}{2\mathrm{v}%
_{0}^{3}}\frac{c^{2}}{\mathcal{E}_{0}\gamma _{0}^{2}}\mathcal{P}^{2}-\left(
Ue^{i\psi }+U^{\ast }e^{-i\psi }\right) -\int d\psi eE_{z},  \label{Hef}
\end{equation}%
where 
\begin{equation}
\Delta =\frac{\omega }{c}+sq-\left( \omega -\omega _{0}\right) \frac{1}{%
\mathrm{v}_{0}}  \label{det}
\end{equation}%
is the resonant detuning and 
\begin{equation}
U\left( z\right) =\frac{e\theta _{s}e^{-\kappa b}E_{0}\left( z\right) }{2i}.
\label{uz}
\end{equation}%
The Vlasov equation 
\begin{equation}
\frac{\partial F}{\partial z}+\frac{\partial H}{\partial \mathcal{P}}\frac{%
\partial F}{\partial \psi }-\frac{\partial H}{\partial \psi }\frac{\partial F%
}{\partial \mathcal{P}}=0  \label{Vl}
\end{equation}%
for the phase-space distribution function $F\left( z,\psi ,\mathcal{P}%
\right) $ will be 
\begin{equation*}
\frac{\partial F}{\partial z}+\left( \Delta +\left( \omega -\omega
_{0}\right) \frac{1}{\mathrm{v}_{0}^{3}}\frac{c^{2}}{\mathcal{E}_{0}\gamma
_{0}^{2}}\mathcal{P}\right) \frac{\partial F}{\partial \psi }
\end{equation*}%
\begin{equation}
+\left( \left( 
\frac{e\theta _{s}e^{-\kappa b}E_{0}\left( z\right) e^{i\psi }}{2}%
+c.c.\right) +eE_{z}\right) \frac{\partial F}{\partial \mathcal{P}}=0.
\label{Vl1}
\end{equation}%

The Maxwell equations for the slowly varying envelope $E_{0}\left( z\right) $
of the probe wave and space charge field can be written as%
\begin{equation}
\frac{\partial E_{0}}{\partial z}+\frac{\mu }{2}E_{0}=-\pi e\frac{\theta
_{s}e^{-\kappa b}}{c}\overline{e^{-i\psi }\int \mathrm{v}FdP},  \label{Max1}
\end{equation}%
\begin{equation}
E_{z}=-i\frac{4\pi e}{\omega }\overline{e^{-i\psi }\int \mathrm{v}FdP},
\label{Max2}
\end{equation}%
where bar denotes averaging over space and time much larger than ($k^{-1}$, $%
\omega ^{-1}$). To take into account the probe wave attenuation, in Eq. (\ref%
{Max1}) we have introduced the absorption coefficient $\mu $. The obtained
Eqs. (\ref{Vl1}), (\ref{Max1}) and (\ref{Max2}) are the self-consistent set
of equations for the considered graphene plasmon-based FEL.

\section{High gain regime of lasing for a cold electron beam}

For the given field $E_{0}$ and initial distribution function the
self-consistent set of equations (\ref{Vl1}), (\ref{Max1}), and (\ref{Max2})
describe generation and amplification processes. Next, we will investigate
collective instability which causes the growth of the initial seed. For the
electron beam distribution function we seek the solution as 
\begin{equation}
F=F_{0}\left( \mathcal{P}\right) +\left( F_{1}\left( z,\mathcal{P}\right)
e^{i\psi }+\mathrm{c.c.}\right) ,  \label{FF}
\end{equation}%
where $F_{0}(\mathcal{P})=N_{0}\delta (\mathcal{P})$, with the Dirac delta
function $\delta (\mathcal{P})$, $N_{0}$-is the mean density of the beam. We
will assume initially modulated electron beam 
\begin{equation}
F_{1}\left( z=0,\mathcal{P}\right) =\eta N_{0}\delta (\mathcal{P}),
\label{mod}
\end{equation}%
where $\eta <1$ is the modulation depth. For the downconversion, because of
long-wavelength $\lambda $ of the output radiation the modulation can be of
Dicke nature: $N_{0}^{-1/3}\lesssim \lambda $. For upconversion, $%
N_{0}^{-1/3}>>\lambda $ we need the special modulation stage, which can be
done by quantum modulation \cite{Avetissian-Book} of the electron beam \cite%
{QM1,QM2,QM3} using femtosecond-pulsed lasers. At the exact resonance $%
\Delta =0$, for the slowly varying envelope function $E_{0}\left( z\right) $
from Eqs. (\ref{Vl1}), (\ref{Max1}), and (\ref{Max2}) one can obtain the
integra-differential equation%
\begin{equation}
\frac{\partial E_{0}}{\partial \widetilde{z}}+\frac{\mu }{2\Gamma }E_{0}=-%
\frac{e\pi \theta _{s}\mathrm{\beta }_{0}e^{-\kappa b}}{\Gamma }\eta
N_{0}+\int\limits_{0}^{\widetilde{z}}\left( \widetilde{z}-\widetilde{z}%
^{\prime }\right) \left[ iE_{0}-\Lambda _{p}^{2}\left( \frac{\partial E_{0}}{%
\partial \widetilde{z}^{\prime }}+\frac{\mu }{2\Gamma }E_{0}\right) \right] d%
\widetilde{z}^{\prime },  \label{Max3}
\end{equation}%
where $\widetilde{z}=z\Gamma $, and 
\begin{equation}
\Gamma =\left( \frac{\left\vert \omega -\omega _{0}\right\vert }{c}\frac{\pi
r_{e}\theta _{s}^{2}N_{0}}{2\gamma _{0}^{3}\mathrm{\beta }_{0}^{2}}%
e^{-2\kappa b}\right) ^{1/3}  \label{gain}
\end{equation}%
is the gain factor, $r_{e}$ is the electron classical radius, $\Lambda _{p}=%
\sqrt{8c\Gamma /\omega \theta _{s}^{2}}$ is the space charge parameter.
Taking into account Eqs. (\ref{intpar}) and (\ref{res_freq}), the gain
factor $\Gamma $ can be explicitly written as%
\begin{equation}
\Gamma =\left( \frac{\omega _{0}}{c}\left\vert 1+sn\left( \omega _{0}\right)
\right\vert \left( 1+\mathrm{\beta }_{0}\right) \left( \frac{1\mathbf{+}%
\frac{s\mathrm{\beta }_{0}}{n\left( \omega _{0}\right) }}{1+sn\left( \omega
_{0}\right) \mathrm{\beta }_{0}}\right) ^{2}\frac{\xi _{0p}^{2}\pi r_{e}N_{0}%
}{2\gamma _{0}^{3}\mathrm{\beta }_{0}^{3}}e^{-2\kappa b}\right) ^{1/3}.
\label{gain2}
\end{equation}%
where $\xi _{0p}=eE_{0p}/(mc\omega _{0})\ $is the GP wave and electron
interaction parameter. 
\begin{figure}[tbp]
\includegraphics[width=.96\textwidth]{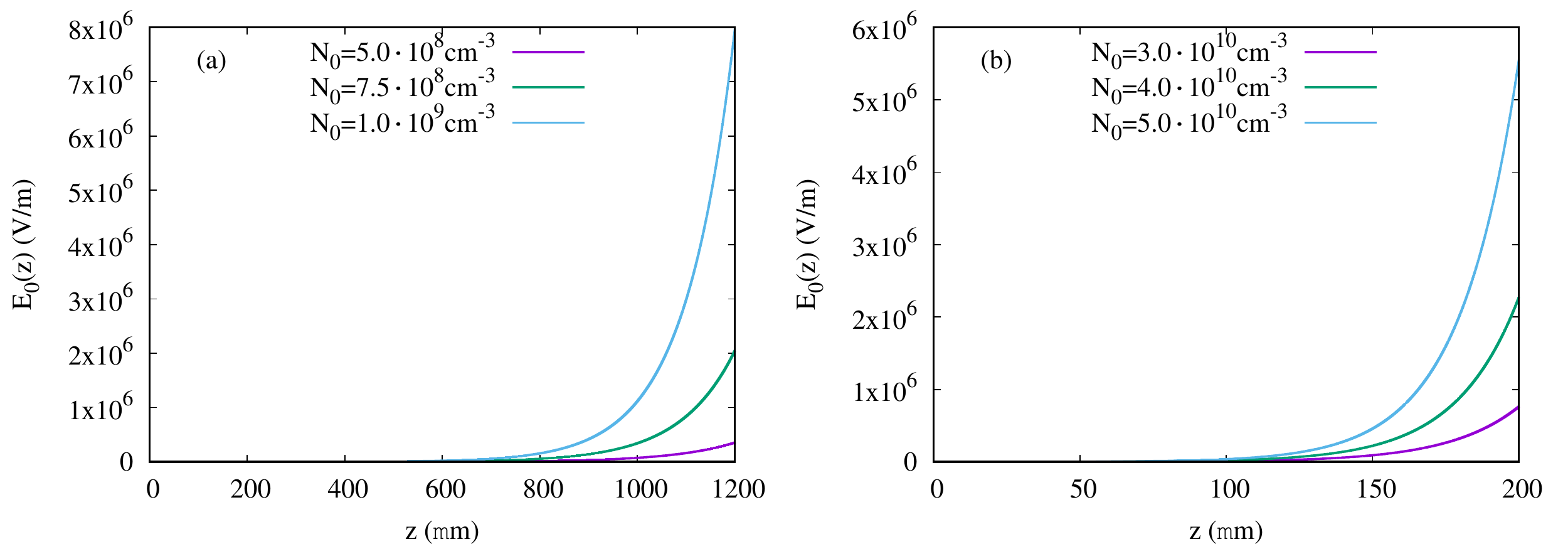}
\caption{ Downconversion at the Cherenkov resonance. We assume Cherenkov
resonance detuning $\protect\delta =1-n\left( \protect\omega _{0}\right) 
\mathrm{\protect\beta }_{0}\simeq 0.1$, modulation depth $\protect\eta =0.1$%
, GP wave field amplitude $E_{0p}=3\ \mathrm{GV/m}$\textrm{.} (a): $\protect%
\lambda _{0}=1.5\ \mathrm{\protect\mu m}$, $E_{F}=0.65\ \mathrm{eV}$, $%
n\left( \protect\omega _{0}\right) \simeq 200$, electron kinetic energy $m%
\mathrm{v}_{0}^{2}/2=5.17\ \mathrm{eV}$\textrm{,} and (b): $\protect\lambda %
_{0}=12\ \mathrm{\protect\mu m}$, $E_{F}=0.12\ \mathrm{eV}$, $n\left( 
\protect\omega _{0}\right) \simeq 90$, electron kinetic energy $m\mathrm{v}%
_{0}^{2}/2=25.5\ \mathrm{eV}$.}
\end{figure}
\begin{figure}[tbp]
\includegraphics[width=.96\textwidth]{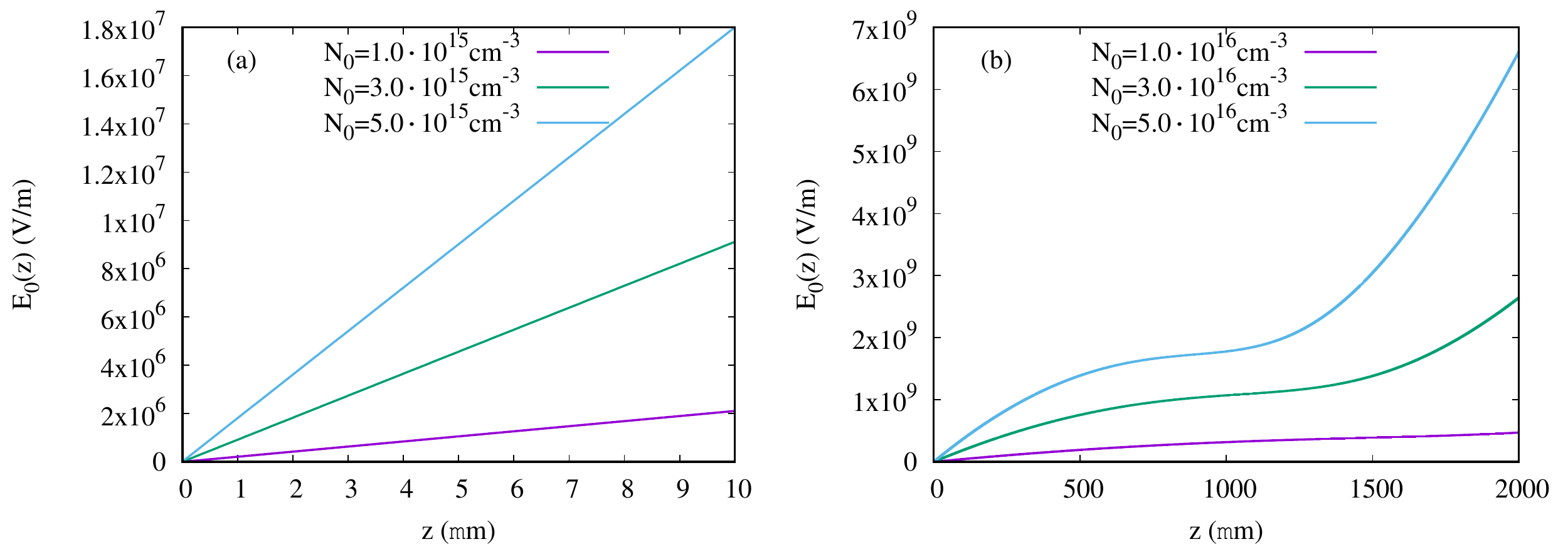}
\caption{Upconversion at the backscattering of GPs with free wavelength $%
\protect\lambda _{0}=30\ \mathrm{\protect\mu m}$ on the electron beam with
Lorentz factor $\protect\gamma _{0}=10$. For the graphene we assume $%
E_{F}=0.34\ \mathrm{eV}$\textrm{, }$n\left( \protect\omega _{0}\right)
\simeq 10$, and GP wave field amplitude $E_{0p}=1\ \mathrm{GV/cm}$. (a)
Superradiant linear growth regime with modulation depth $\protect\eta %
=10^{-1}$. (b) Exponential growth regime with $\protect\eta =10^{-2}$.}
\end{figure}
We will consider a tight confinement regime $n\left( \omega _{0}\right) >>1$%
. For upconversion, it is of interest a relativistic electron beam and
counterpropagating ($s=1$) GP wave. From the expression for resonant
frequency (\ref{res_freq}), we have 
\begin{equation}
\omega \simeq 2\gamma _{0}^{2}n\left( \omega _{0}\right) \omega _{0}
\label{ww}
\end{equation}%
and the gain factor is approximated as 
\begin{equation}
\Gamma \simeq \left( \frac{\omega _{0}}{c}\frac{\xi _{0p}^{2}\pi r_{e}N_{0}}{%
\gamma _{0}^{3}n\left( \omega _{0}\right) }e^{-2\kappa b}\right) ^{1/3}.
\label{gain3}
\end{equation}%
For downconversion, it is of interest nonrelativistic electron beam with a
velocity close to Cherenkov resonance $\left\vert \delta \right\vert
=\left\vert 1-n\left( \omega _{0}\right) \mathrm{\beta }_{0}\right\vert <<1$%
. In this case, we have $\omega =\omega _{0}\left\vert \delta \right\vert $
and the gain factor is given as%
\begin{equation}
\Gamma \simeq \left( \frac{\omega _{0}}{c}\frac{n^{2}\left( \omega
_{0}\right) }{\delta ^{2}}\frac{\xi _{0p}^{2}\pi r_{e}N_{0}}{2\mathrm{\beta }%
_{0}^{3}}e^{-2\kappa b}\right) ^{1/3}.  \label{gain4}
\end{equation}%
As is seen, in this case we have several enhancement factors compared with
the relativistic one: $n^{2}\left( \omega _{0}\right) >>1$, $\mathrm{\beta }%
_{0}^{-3}>>1$, and $\delta ^{-2}>>1$. The latter allows to use dilute
electron beams.

To get an intuitive understanding of the generation process, we will
explicitly write down the solution in extreme cases. When $\mu <<\Gamma $
and $\Lambda _{p}<<1$, we have the exponential growth $E_{0}\sim e^{\sqrt{3}%
\Gamma z/2}$ of the probe wave. For the low beam currents $\Gamma \sim \mu
,\Lambda _{p}<<1$ and at short distances $z<<\mu ^{-1}$, we have linear
growth 
\begin{equation}
E_{0}\left( z\right) \simeq e\pi \theta _{s}\mathrm{\beta }_{0}\eta
N_{0}e^{-\kappa b}z  \label{sup}
\end{equation}%
of superradiant nature that at the long distances is saturated to value 
\begin{equation}
E_{0}\left( z\right) \simeq \frac{2e\pi \theta _{s}\mathrm{\beta }_{0}\eta
N_{0}e^{-\kappa b}}{\mu }.  \label{sat}
\end{equation}%
In general case, we have solved Eq. (\ref{Max3}) numerically considering two
setups for downconversion and two -for upconversion. For both regimes, we
assume strong coupling limit $e^{-\kappa b}\simeq 1$. In Fig. 2, we present
the results for downconversion at the Cherenkov resonance. We assume a
detuning for Cherenkov resonance to be $\delta =1-n\left( \omega _{0}\right) 
\mathrm{\beta }_{0}\simeq 0.1$, modulation depth $\eta =0.1$, and GP wave
field amplitude $E_{0p}=3\ \mathrm{GV/m}$\textrm{.} For all electron beam
densities we assume an absorption coefficient $\mu =0.2\Gamma $. In Fig.
2(a), we take $\lambda _{0}=2\pi c/\omega _{0}=1.5\ \mathrm{\mu m}$, $%
E_{F}=0.65\ \mathrm{eV}$, which (according to Eq. (\ref{sg})) for the
confinement factor gives $n\left( \omega _{0}\right) \simeq 200$. For such a
value of $n\left( \omega _{0}\right) $ one needs electrons with kinetic
energy $m\mathrm{v}_{0}^{2}/2=5.17\ \mathrm{eV}$\textrm{,} and at the
relatively dilute electron beam we have the exponential growth regime. In
Fig 2(b), we take $\lambda _{0}=12\ \mathrm{\mu m}$, $E_{F}=0.12\ \mathrm{eV}
$. For these values from Eq. (\ref{sg}) we get $n\left( \omega _{0}\right)
\simeq 90$, which in turn demands resonant value for electron kinetic energy 
$m\mathrm{v}_{0}^{2}/2=25.5\ \mathrm{eV}$. To achieve an exponential growth
regime in this case we need an electron beam of one order large density.

In Fig. 3, we present the solution of Eq. (\ref{Max3}) for upconversion at
the backscattering of GPs with free wavelength $\lambda _{0}=30\ \mathrm{\mu
m}$ on the electron beam with Lorentz factor $\gamma _{0}=10$. For a
graphene, we assume Fermi energy $E_{F}=0.34\ \mathrm{eV}$\textrm{, }which
gives\textrm{\ }$n\left( \omega _{0}\right) \simeq 10$. As is clear from Eq.
(\ref{gain3}), for the reasonable values of the gain factor we need a strong
GP wave field $E_{0p}=1\ \mathrm{GV/cm}$ and dense electron beam with $%
N_{0}\sim 10^{15}-10^{16}$. At that, we have the upconversion (\ref{ww}) of
mid-infrared to extreme UV radiation: $\lambda \simeq 15\ \mathrm{nm}$. For
all densities we assume that $\mu =\Gamma $. In Fig. 3(a), we consider short
interaction distance where we have linear growth (\ref{sup}) and
superradiation $E_{0}^{2}\sim N_{0}^{2}$. In Fig. 3(b), we see an
exponential growth regime.

\section{Conclusion}

We have developed a new mechanism for the tabletop nano-FEL in graphene
based on the coherent scattering of free electrons on plasmons. The
analytical description of the issue in the scope of self-consistent theory
within the plasma dynamics in a 2D nanostructure and EM radiation field has
been done. For the ultimate quantitative results also numerical calculation
for the generation process has been made. The proposed nano-FEL scheme is
capable to achieve lasing from THz to extreme UV domain due to the
stimulated scattering of graphene plasmons on free electrons. We have
considered downconversion as well as upconversion. It has been shown that
the coherent downconversion of infrared to THz radiation can be achieved
using a source of electrons of several eV resonantly coupled with graphene
plasmons, which can be implemented in on-chip configurations. It has been
shown that due to strongly confined graphene plasmons, the upconversion of
mid-infrared to extreme UV radiation can be achieved with mildly
relativistic electron beams which is the promising mechanism of tabletop
short wavelength nanolaser. For short interaction distances, one can use a
superradiant regime of the generation with coherently shaped electron
batches.

\textit{Acknowledgement}.---This work was supported by the RA State Committee of Science and Belarusian
Republican Foundation for Fundamental Research (RB) in the frame of the
joint research project SCS 18BL-020.

\end{document}